\documentclass[a4paper,fleqn]{cas-dc}
\usepackage{xurl}
\usepackage{threeparttablex}
\usepackage{booktabs}

\usepackage[numbers]{natbib}
\usepackage{subcaption}
\usepackage{caption}
\usepackage{multirow}
\usepackage{stfloats}
\usepackage{float}
\usepackage{booktabs}
\usepackage{threeparttable}
\usepackage{adjustbox}
\usepackage{placeins}
\usepackage{capt-of}

\def\tsc#1{\csdef{#1}{\textsc{\lowercase{#1}}\xspace}}
\tsc{WGM}
\tsc{QE}

\begin{document}
\let\WriteBookmarks\relax
\def\floatpagepagefraction{1}
\def\textpagefraction{.001}  

\shortauthors{Talavera-Mu{\~n}oz et al.}


\title [mode = title]{A four-step validation protocol for radiomic survival analysis in small cohorts: a body composition case study}

\shorttitle{Validation protocol for radiomic survival analysis}



%

\author[1]{Mariela Talavera-Mu{\~n}oz}[orcid=0009-0004-8980-5434]
\cormark[1] 
\ead{mariela.talavera@estudiante.uam.es}

\author[1]{Cecilia Diana-Albelda}

\author[2, 3]{Miguel Antonio Sampedro-Nu{\~n}ez}

\author[3, 4]{Victor Rodriguez-Laval}
 
\author[1]{Jesus Bescos}

\author[1]{{\'A}lvaro Garc{\'i}a-Mart{\'i}n}







\affiliation[1]{organization={Video Processing and Understanding Lab},
            addressline={Escuela Polit{\'e}cnica Superior, Universidad Aut{\'o}noma de Madrid, Calle Francisco Tom{\'a}s y Valiente 11},
            city={Madrid},
            postcode={28049},
            state={Madrid},
            country={Spain}}

\affiliation[2]{organization={Department of Endocrinology and Nutrition, Hospital Universitario de La Princesa, Instituto de Investigaci{\'o}n Sanitaria de La Princesa},
            addressline={Calle de Diego de Le{\'o}n 62},
            city={Madrid},
            postcode={28006},
            state={Madrid},
            country={Spain}}

\affiliation[3]{organization={Instituto de Investigaci{\'o}n Sanitaria del Hospital Universitario de La Princesa},
            addressline={Calle de Diego de Le{\'o}n 62},
            city={Madrid},
            postcode={28006},
            state={Madrid},
            country={Spain}}

\affiliation[4]{organization={Department of Radiodiagnostics, Hospital Universitario de La Princesa},
            addressline={Calle de Diego de Le{\'o}n 62},
            city={Madrid},
            postcode={28006},
            state={Madrid},
            country={Spain}}





\cortext[cor1]{Corresponding author}

\begin{abstract}
Gastroenteropancreatic neuroendocrine tumors (GEP-NET) lack prognostic biomarkers, and computed tomography (CT)-based body composition radiomics at the third lumbar vertebra (L3) is a promising candidate. Small cohorts make high-dimensional radiomic models prone to overfitting and single-split evaluations variable and optimistic. 

This methodological study proposes a four-step validation protocol for radiomic survival analysis in small cohorts: (1) false discovery rate (FDR)-corrected feature selection, (2) a dual-path comparison of prior feature selection versus embedded regularization, (3) repeated nested cross-validation (CV) and (4) a negative control with Gaussian noise features. In 78 GEP-NET patients, 535 radiomic features from five L3 compartments were used to evaluate five survival models with Harrell's concordance index ($C$-index), time-dependent area under the curve (AUC) and integrated Brier score (IBS).

Step~1 selected at least one feature in $18\%$ of outer folds under noise (nominal level $20\%$) versus $84\%$ with radiomic features. The dual-path comparison showed no consistent difference between strategies. Repeated nested CV reduced variability across seeds ($41\%$ median reduction in $C$-index standard deviation), but not enough to distinguish models. The nested CV $C$-index exceeded chance by $0.07$--$0.15$ with radiomic features but not clearly under noise ($0.40$--$0.42$ with prior feature selection, $0.47$--$0.55$ with embedded regularization). Frequently selected features tentatively linked greater subcutaneous fat, skeletal muscle compactness and texture to better prognosis.

These findings suggest a modest prognostic signal in L3 radiomics, with considerable uncertainty that the protocol quantifies. The protocol may transfer to other small-cohort radiomic survival studies. Code: \url{https://github.com/marielatalaveram/radiomics-survival-validation-protocol}.
\end{abstract}


\maketitle

\section{Introduction}\label{intro}

\begin{figure*}[!t]
    \centering
    \includegraphics[trim=0cm 7cm 0cm 6cm, clip, width=\textwidth]{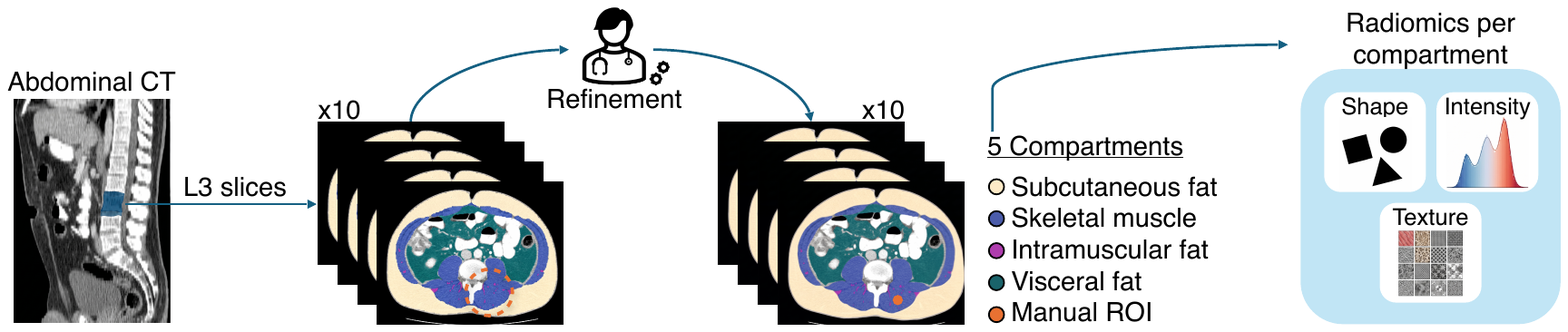}
    \caption{Workflow for radiomic feature extraction. A standardized volume consisting of ten central slices at the L3 vertebral level is selected from the abdominal CT scan. Automatic segmentations are refined by clinical experts to delineate four body composition compartments (skeletal muscle, subcutaneous fat, visceral fat, and intramuscular fat), alongside a fifth compartment, a spherical region of interest (ROI) placed manually in the paraspinal muscles (manual ROI). Finally, 107 radiomic features, encompassing shape, intensity and texture, are extracted from each of the five compartments per patient.}
    \label{fig:data_extraction}
\end{figure*}

Gastroenteropancreatic neuroendocrine tumors (GEP-NET) constitute a rare and biologically heterogeneous group of neoplasms arising from the neuroendocrine cells of the gastrointestinal tract and pancreas. These tumors are characterized by a markedly variable clinical behavior \cite{Dasari2017}: while most patients present with mild and non-specific symptoms following an indolent course over several years, others rapidly develop aggressive disease. Tumor grade, stage, and primary site are routinely used to stratify patients, yet outcomes remain markedly heterogeneous within each risk group \cite{CarmonaBayonas2019}. Molecular markers could help refine risk stratification \cite{rossi2025netest}, but their routine clinical availability is far more limited than that of conventional parameters. There is therefore a need for complementary, non-invasive biomarkers that can be derived from routine diagnostic imaging.

Body composition is a natural candidate. Sarcopenia, the progressive loss of muscle mass and strength \cite{cruzjentoft2019sarcopenia}, has emerged as a potential predictor of mortality in patients with these tumors. However, it captures only one facet of the body composition alterations associated with malnutrition and cachexia, conditions that also affect adipose tissue and are known to worsen GEP-NET prognosis \cite{SebastianValles2022}. Assessing body composition as a whole therefore offers a broader perspective than muscle mass alone, which may be relevant for prognosis.

Several methods exist for measuring body composition. Body mass index, although one of the most widely used nutritional indicators, cannot distinguish fat from lean tissue \cite{Caan2018}, and is therefore insufficient to characterize these compartments.
In contrast, computed tomography (CT) analysis at the third lumbar vertebra (L3) has become the reference standard \cite{SebastianValles2022}: it accurately quantifies muscle and adipose tissue and relies on scans already acquired during routine oncological care \cite{mourtzakis2008practical}.

Conventional L3 analysis, however, is limited to basic cross-sectional metrics. Radiomics offers a way to characterize tissues in much greater depth: by extracting high-dimensional parameters of tissue microarchitecture, it captures the intrinsic spatial heterogeneity of each compartment, providing a promising non-invasive strategy for biomar\-ker development in oncology.
In neuroendocrine tumors, radiomics has been applied to therapy response \cite{Polici2025}, histological grading \cite{Gu2019, Chiti2022}, and prognostic stratification of liver metastases \cite{Martini2020}, but mostly through tumor-derived features. To the best of our knowledge, body composition radiomics remains largely unexplored as a source of prognostic biomarkers in these tumors.

However, addressing this gap is methodologically demanding. GEP-NET are rare \cite{Dasari2017}, so available cohorts are inherently small. This is further aggravated in prognostic studies, which are typically framed as time-to-event (survival) analysis: the effective sample size is given by the number of events (deaths) rather than by the number of patients \cite{peduzzi1995importance}. Radiomics, which yields hundreds of correlated features per patient, is especially vulnerable in this setting, since model validity and reproducibility are difficult to ensure when features far outnumber events \cite{horvat2024radiomics}.

In practice, many studies select features without correction for multiple comparisons \cite{chalkidou2015false}, evaluate performance on a single train/test split \cite{varoquaux2018cross}, and lack negative controls to verify that the identified signal is not an artifact of the analytical pipeline 
\cite{welch2019vulnerabilities, buvat2019dark}. Together with the feature-to-event imbalance, these practices can produce optimistic or highly variable performance
estimates, making it difficult to distinguish genuine signal from noise \cite{varoquaux2018cross}.

To mitigate these shortcomings, reporting and quality-assessment tools such as the Radiomics Quality Score \cite{lambin2017radiomics}, CLEAR \cite{kocak2023clear}, METRICS \cite{kocak2024metrics} and TRIPOD+AI \cite{collins2024tripodai} recommend, among other aspects, appropriate validation and transparent reporting. 
However, they indicate what should be reported or assessed but do not provide a procedure to quantify how validation choices affect performance estimates. The protocol proposed here is intended as a quantitative complement to these guidelines. It combines four controls, each targeting a distinct methodological risk: feature selection with false discovery rate (FDR) correction for multiple comparisons, to limit false discoveries; a dual-path comparison (prior feature selection versus the full set of features with embedded regularization), to assess the sensitivity of the results to the dimensionality-reduction strategy; repeated nested cross-validation (CV), to counter optimistic and highly variable performance estimates; and a negative control, to detect artifacts of the analytical pipeline. As a clinical application, the protocol is used to evaluate the prognostic value of L3 body composition radiomics in GEP-NET across multiple survival models.

\section{Materials and methods}\label{methods}

\subsection{Recruitment and variable collection}

This was a single-center, retrospective study conducted at Hospital Universitario de la Princesa (Madrid, Spain). Patients were drawn from a broader hepatopancreatobiliary oncology database and selected for this study if they met the following criteria: (i) a confirmed diagnosis of GEP-NET; (ii) clinical follow-up at the center; and (iii) availability of a baseline diagnostic CT scan.
Patients were excluded if they had a history of other gastrointestinal malignancies (including hepatocellular carcinoma), incomplete survival records, or imaging studies affected by significant motion artifact, absence of intravenous contrast or image misregistration. All data were pseudonymized prior to analysis.

For each patient, the collected data included survival outcomes, demographic variables, and radiomic features. 
Survival outcomes were defined by two parameters: event status (a binary indicator where 1 denotes death and 0 denotes a right-censored observation, representing patients who did not experience the event during follow-up), and overall survival time (measured in days from diagnosis to the date of death or last clinical contact) \cite{clark2003survival}. 
Patient demographics, specifically age at diagnosis and sex, were used for descriptive and univariate survival analyses.
In contrast, the radiomic features were the only predictors included in the survival models. As shown in Figure~\ref{fig:data_extraction}, the extraction process began with whole-body automatic segmentation using TotalSegmentator \cite{wasserthal2023totalsegmentator}, from which only the L3 region was retained. To ensure 3D homogeneity across the cohort, a standardized axial volume consisting of the ten central slices at the L3 level, corresponding to the minimum slice count available among all subjects, was selected. Automatic segmentations were subsequently reviewed and manually refined by three clinical experts \cite{decaux2023semiautomatic}. 

Within this final volume, five compartments were defined: four body composition compartments (skeletal muscle, subcutaneous fat, visceral fat, and intramuscular fat) \cite{lenchik2019automated} and a fifth compartment, a spherical region of interest placed manually in the paraspinal muscles (manual ROI). Using PyRadiomics \cite{vanGriethuysen2017computational}, 107 radiomic features were extracted from each of the five compartments, totaling 535 features per patient.

\subsection{Proposed validation protocol}\label{protocol}

\begin{figure*}[!t]
    \centering
    \includegraphics[trim=1cm 5cm 1cm 4.8cm, clip, width=\textwidth]{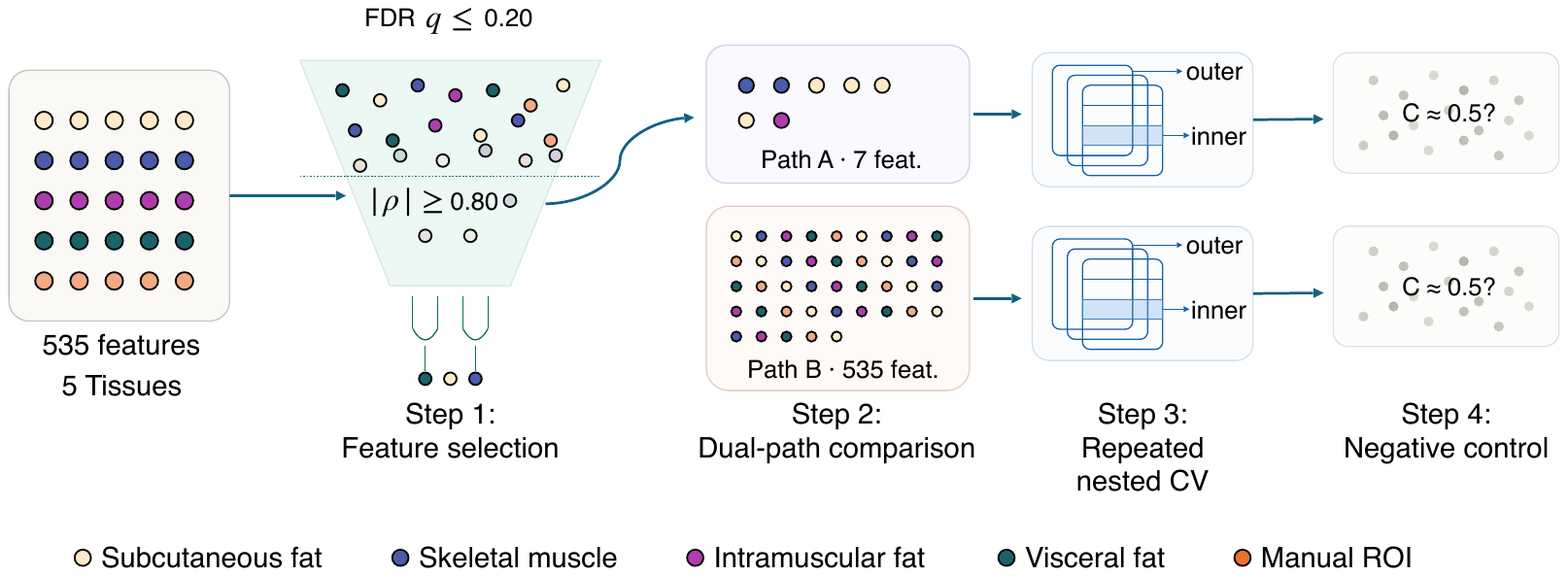}
    \caption{Proposed validation protocol for radiomic survival analysis. Step~1 selects features from the 535 radiomic features extracted across five compartments: a false discovery rate (FDR) filter (univariate Cox analysis with FDR correction, $q \leq 0.20$) followed by a redundancy filter (correlation clustering, $|\rho| \geq 0.80$). Step~2 establishes a dual-path comparison of dimensionality-reduction strategies: Path~A fits the survival models on the selected set of features, whereas Path~B fits them on the full set of 535 features, relying on embedded regularization. Step~3 evaluates both paths with repeated nested cross-validation (CV): 5-fold CV stratified by event status, with hyperparameters tuned in an inner loop, repeated with ten random seeds to quantify the variability of the estimates. Finally, Step4 applies the analytical pipeline (Steps 1--3) to Gaussian noise features as a negative control.}
    \label{fig:validation_protocol}
\end{figure*}

The proposed validation protocol, illustrated in Figure~\ref{fig:validation_protocol}, comprises four steps: (1) FDR-corrected feature selection; (2) a dual-path comparison, in which five survival model configurations are fitted either on the selected set of features (Path~A, prior feature selection) or on the full set of 535 features (Path~B, embedded regularization); (3) repeated nested CV of both paths; and (4) a negative control on Gaussian noise features.

\subsubsection{Feature selection}\label{sec:featureselection}

Feature selection (Step~1) comprised two stages. In the first stage (FDR filter), a univariate Cox analysis based on the proportional hazards (PH) model \cite{cox1972regression, hosmer2008applied} was performed for each feature, with Benjamini--Hochberg FDR correction \cite{benjamini1995controlling} to account for multiple comparisons. A permissive threshold (FDR $q \leq 0.20$) was adopted \cite{subramanian2005gene}, as stricter thresholds selected too few features in the study cohort (0 at $q\leq 0.05$; 2 at $q \leq 0.10$). This threshold was used as an exploratory screening criterion rather than to establish statistical significance.

In the second stage (redundancy filter) \cite{bogowicz2017computed}, the features selected by the FDR filter were grouped by Spearman correlation-based hierarchical clustering (single linkage), so that features linked by $|\rho| \geq 0.80$, directly or through other features, formed a single cluster. Within each cluster, a single representative feature was selected based on the highest Harrell's concordance index ($C$-index) \cite{harrell1996multivariable} obtained from the univariate Cox analysis \cite{leger2017comparative}. The features that passed both stages formed the selected set of features used by Path~A.

\subsubsection{Dual-path comparison: survival models}
\label{survmodels}

The second step compares the two paths to examine whether performance depends on the dimensionality-reduction strategy. Five survival model configurations were considered: Cox regression \cite{cox1972regression} with Ridge \cite{hoerl1970ridge} or LASSO \cite{tibshirani1997lasso} penalization (Cox Ridge and Cox LASSO), random survival forest (RSF) \cite{ishwaran2008random}, gradient boosting survival analysis (GBSA) \cite{friedman2001greedy, hothorn2006survival}, and survival support vector machine (SSVM) \cite{vanbelle2011support, shivaswamy2007support}. Each configuration was fitted independently on the selected set of features from Step~1 (Path~A) and on the full set of features, relying on embedded regularization (Path~B), yielding ten model--path combinations.
Hyperparameters of each configuration were tuned by grid search within the nested CV.

\paragraph{Cox regression.}

Cox regression models were fitted with Ridge ($\ell_2$-norm) and LASSO ($\ell_1$-norm) penalization \cite{simon2011regularization}. The grid for the penalty strength $\lambda$ comprised ten log-spaced values in the range $[10^{-1}, 10^{2}]$.

\paragraph{Ensemble-based survival models.}

Two tree-based ensemble algorithms were considered. 

The first, RSF, used a grid covering the number of trees ($\{500, 1000\}$), the minimum number of samples required to split an internal node ($\{15, 20, 25\}$), the minimum number of samples per terminal leaf ($\{15, 20, 25, 30\}$), and the number of features considered at each split. 

The second, GBSA, was built on shallow decision trees; to control overfitting, its grid covered the learning rate ($\eta \in \{0.01, 0.05, 0.1, 0.5\}$), the subsampling ratio ($\{0.5, 0.7, 1.0\}$), and the maximum tree depth ($\{1, 2\}$).

\paragraph{Survival support vector machine (SSVM).}

A kernel-based SSVM was implemented in regression mode. The grid covered three kernel functions (linear, polynomial, and radial basis function), the penalty strength $\lambda \in [10^{-4}, 10^{2}]$ on a logarithmic scale, and, for the non-linear kernels, the kernel coefficient $\gamma \in \{1/p, 10^{-3}, 10^{-2}, 10^{-1}, 1\}$, where $p$ is the number of features. 

\subsubsection{Evaluation} 

The final two steps of the proposed protocol address complementary questions: repeated nested CV provides a less variable performance estimate, whereas the negative control assesses whether the analytical pipeline yields above-chance performance in the absence of true signal.

\paragraph{Repeated nested cross-validation.}

As the third step of the protocol, Steps~1 (feature selection) and 2 (dual-path comparison) were embedded within a 5-fold nested CV stratified by event status, repeated with ten random seeds (42--51). In turn, each outer fold was held out for evaluation (held-out fold), and the remaining four folds (outer-training data) were used exclusively for preprocessing, feature selection, hyperparameter tuning, and model fitting; all ten model--path combinations were then evaluated on the held-out fold. Hyperparameters were tuned by grid search in an inner loop on the outer-training data, selecting the value that maximized the $C$-index. The inner loop used 5-fold CV for the Cox models, and Monte Carlo CV \cite{xu2001monte} with five random splits for SSVM, RSF, and GBSA, each split holding out $30\%$ of the outer-training data as validation data.

Performance was evaluated with two discrimination metrics, the $C$-index and the time-dependent area under the receiver operating characteristic curve (AUC) \cite{heagerty2000time}, for which higher values indicate better performance, and with the integrated Brier score (IBS) \cite{graf1999assessment}. Unlike the $C$-index and AUC, the IBS measures overall prediction error, combining discrimination and calibration, so lower values indicate better performance; it is not defined for SSVM. The reference value is $0.25$, obtained by a model that assigns a $50\%$ survival probability to every patient at every time point \cite{graf1999assessment, steyerberg2019clinical}.
If Step~1 selected no features in the outer-training data of a given outer fold, the corresponding evaluation was assigned chance-level discrimination ($C$-index and AUC of $0.5$), with the IBS treated as undefined. For each seed, performance was summarized as the mean across the five outer folds, and results are reported as the mean $\pm$ standard deviation (SD) between seeds.

Path~A and Path~B were compared on the same outer folds of the nested CV. For each seed, the paired difference between paths was computed fold by fold and averaged across the outer folds. The comparison was summarized with two measures: $\Delta_{A-B}$, the mean paired difference across seeds, and $n_A$, the number of seeds in which Path~A performed better.

\paragraph{Negative control.}
As the fourth and final step, the analytical pipeline (Steps~1--3) was applied after replacing the radiomic features with Gaussian noise features of the same dimensions, keeping the survival outcomes unchanged.

\subsection{Validation scheme for performance estimates}

To examine how the choice of validation scheme affects performance estimates, the same analytical pipeline was applied under two schemes: (i) a single 70/30 train/test split stratified by event status (single split), with $70\%$ of the study cohort as training data and $30\%$ as test data, and (ii) the repeated nested CV of Step~3. For each of the ten seeds, both schemes were applied to the same cohort, and all other steps were identical and fitted exclusively on the training data (the outer-training data in the nested CV): $z$-score standardization, feature selection, and hyperparameter tuning in an inner loop \cite{demircioglu2021measuring, varma2006bias}.

The discrepancy between schemes was quantified, for each model--path combination and metric, with two measures: the mean difference ($\Delta$) as the mean across seeds of the paired difference between the single-split and nested CV estimates; and the mean absolute difference ($|\Delta|$), which indicates the typical size of the discrepancy regardless of direction. For the IBS, the sign of $\Delta$ was inverted so that, for all metrics, positive values indicate a more favorable estimate under the single split.

\subsection{Univariate survival analysis}

In addition to the protocol, a descriptive univariate survival analysis was performed after the analytical pipeline. This analysis was not used for feature selection or model evaluation. Demographic variables (age and sex) were analyzed in the study cohort, and the radiomic features selected in Step~1 were analyzed in the training data of the single split.

For each variable, a univariate Cox analysis was performed to estimate the hazard ratio (HR, expressed per SD for continuous variables), its 95\% confidence interval (CI), and the $C$-index. For the radiomic features, statistical significance was assessed with the Wald test, with the Benjamini--Hochberg FDR correction for multiple comparisons applied to the 535 features, as in Step~1. For the demographic variables, which were treated as categorical (age was grouped as $<$55, 55--65, 65--75, and $\geq$75 years), significance was assessed by comparing Kaplan--Meier (KM) survival curves with the log-rank test. The PH assumption was tested with scaled Schoenfeld residuals \cite{grambsch1994proportional}.

To evaluate the consistency of selection, Step~1 was repeated on the training data of 100 random 70/30 splits stratified by event status. The selection frequency (SF) of each feature was defined as the proportion of repetitions in which it was selected, and the conditional selection frequency (CSF) as the same proportion restricted to repetitions in which at least one feature was selected. Features with lower SF and CSF were interpreted with greater caution. This analysis was exploratory and did not alter the selected set of features.

\section{Results}\label{results}

\subsection{Cohort characteristics}

The initial cohort comprised 113 patients. After applying the exclusion criteria, 78 patients ($69\%$) were retained for the final analysis (Table~\ref{tab:demographics}). The study cohort was balanced by sex ($48.7\%$ female, $51.3\%$ male), with a median age at diagnosis of 63 years and a median observed overall survival time of 2,523 days (events and censored patients combined). During follow-up, 27 patients ($34.6\%$) died, while the remaining were censored (Table~\ref{tab:demographics}). For subsequent survival analysis, age at diagnosis was grouped into four groups of similar size ($<$55, 55--65, 65--75, $\geq$75 years).

\begin{table*}[!t]
    \centering
    \footnotesize
    \caption{Demographic variables of the study cohort (N=78), stratified by survival outcome. Event denotes patients who died, whereas censored refers to patients who did not experience the event during the follow-up period. Overall survival time is the observed time from diagnosis to death or last clinical contact, so censored patients contribute their follow-up time. Age groups are defined at diagnosis. Values are presented as counts (\%) or medians (interquartile range, IQR).}
    \label{tab:demographics}
    \begin{tabular}{l c c c}
        \toprule
         & \textbf{Overall} & \textbf{Events} & \textbf{Censored} \\
        \midrule
        Study cohort, N (\%) & 78 (100.0\%) & 27 (34.6\%) & 51 (65.4\%) \\
        \midrule
        Sex, n (\%) & & & \\
        \hspace{1em}Female & 38 (48.7\%) & 16 (59.3\%) & 22 (43.1\%) \\
        \hspace{1em}Male & 40 (51.3\%) & 11 (40.7\%) & 29 (56.9\%) \\
        \midrule
        Age (years), median (IQR) & 63.0 (52.2--75.0) & 72.0 (62.0--77.5) & 58.0 (46.0--73.0) \\
        \midrule
        Age groups (years), n (\%) & & & \\
        \hspace{1em}$<55$ & 21 (26.9\%) & 3 (11.1\%) & 18 (35.3\%) \\
        \hspace{1em}55--65 & 22 (28.2\%) & 5 (18.5\%) & 17 (33.3\%) \\
        \hspace{1em}65--75 & 19 (24.4\%) & 11 (40.7\%) & 8 (15.7\%) \\
        \hspace{1em}$\geq75$ & 16 (20.5\%) & 8 (29.6\%) & 8 (15.7\%) \\
        \midrule
        Overall survival time (days), median (IQR) & 2523.0 (1671.0--3937.0) & 1704.0 (884.5--3494.5) & 2627.0 (2028.0--4348.5) \\
        \bottomrule
    \end{tabular}
\end{table*}

\subsection{Univariate survival analysis}

\begin{figure*}[!ht]
    \centering
    \begin{subfigure}[b]{0.42\textwidth}
        \centering
        \includegraphics[width=\textwidth]{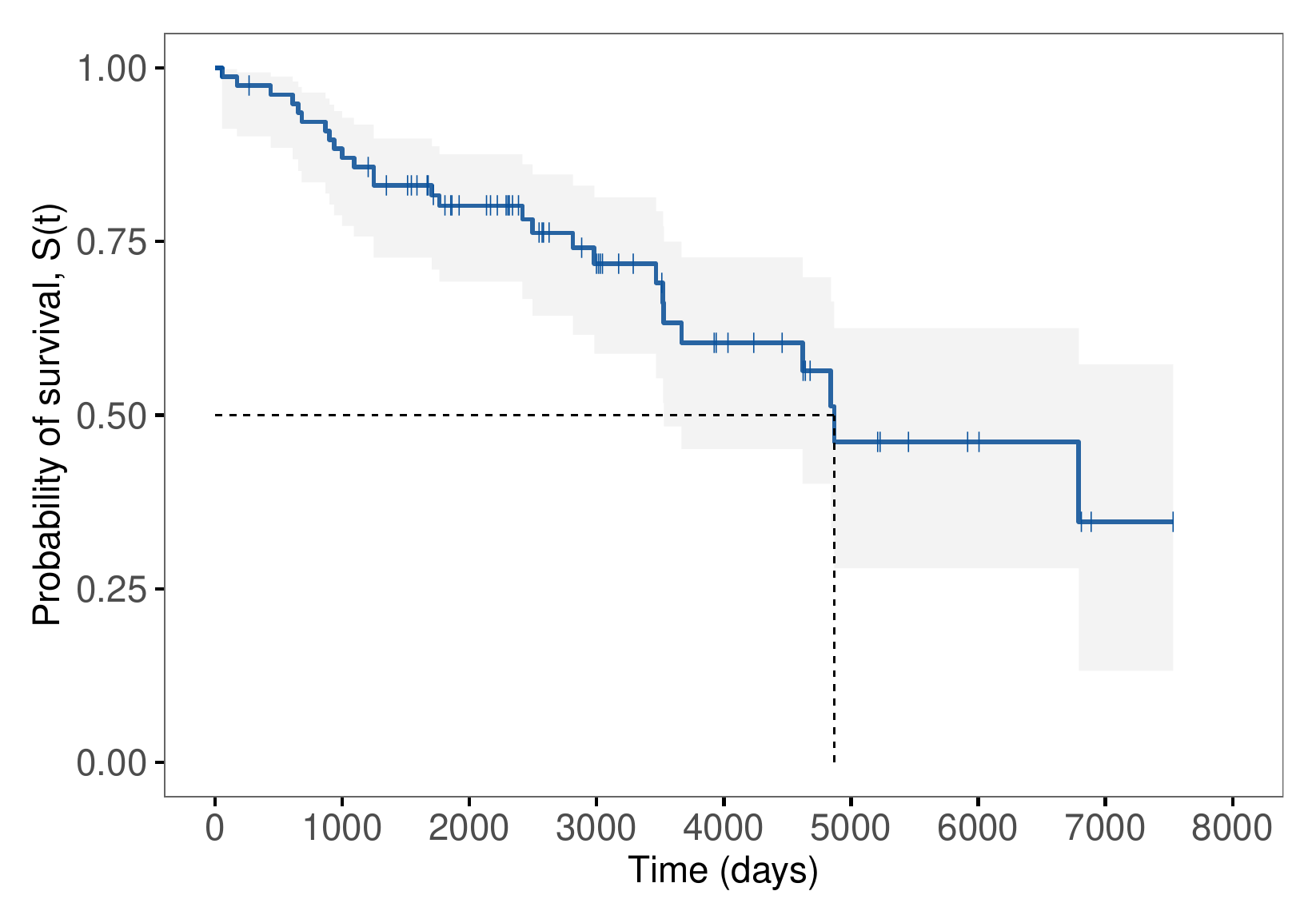}
        \caption{Overall survival curve}
        \label{fig:km-general}
    \end{subfigure}
    \hfill  
    \begin{subfigure}[b]{0.42\textwidth}
        \centering
        \includegraphics[width=\textwidth]{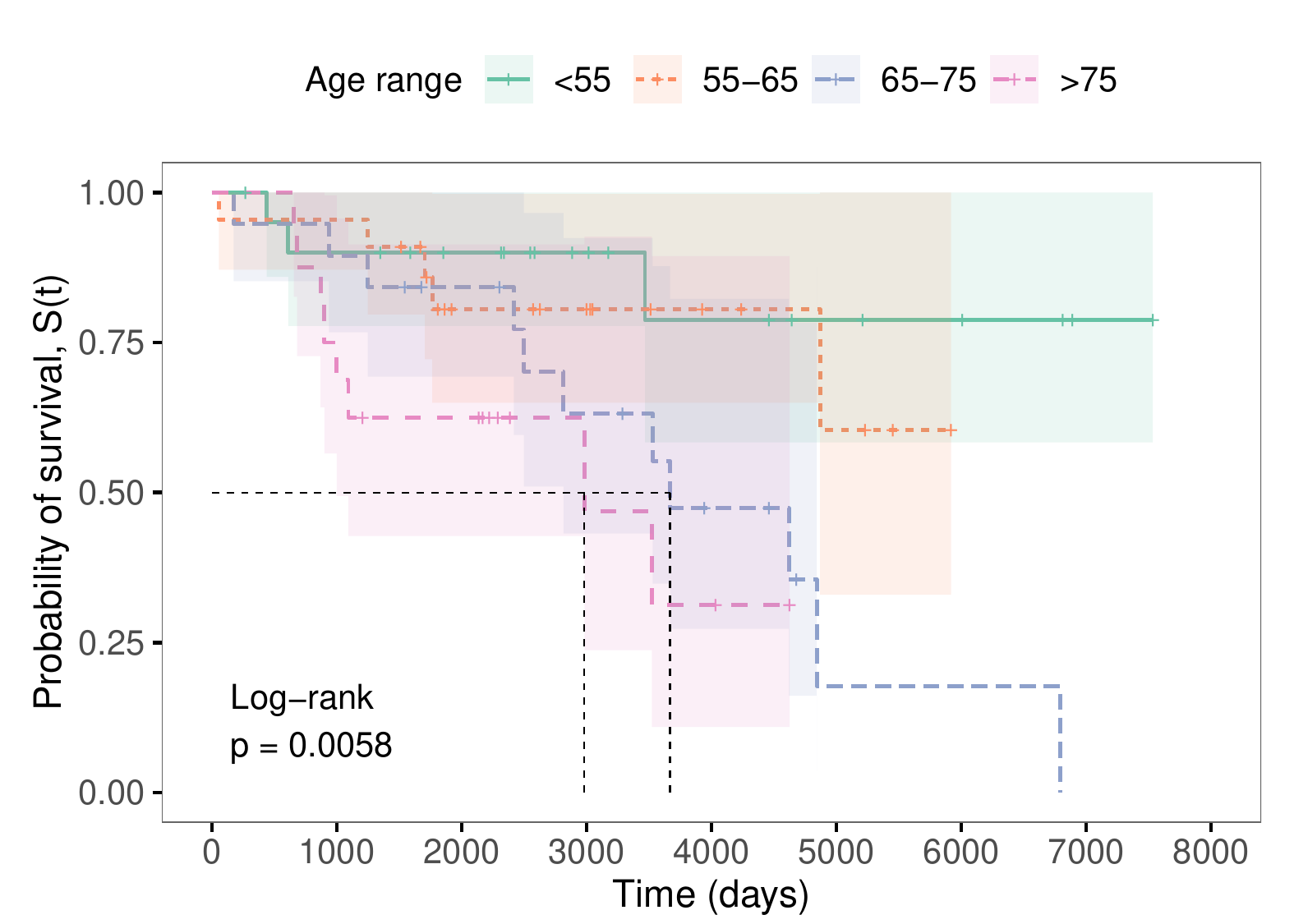}
        \caption{Survival curve by age group}
        \label{fig:km-age}
    \end{subfigure}
    
    \caption{Kaplan--Meier survival curves for (a) the study cohort and (b) patients stratified by age group. Follow-up time is expressed in days, and survival probability is denoted as $S(t)\in [0, 1]$. Age groups in (b) are color-coded: the youngest group ($<$55) is plotted in green, the 55--65 group in orange, 65--75 in blue, and $\geq75$ group in pink.}

    \label{fig:km-combined}
\end{figure*}

KM analysis (Figure~\ref{fig:km-general}) showed a gradual decline in survival over the first 3,000 days, with survival probability remaining above $70\%$. A steeper decline followed, with median survival reached at approximately 4,850 days. Beyond this point the curve flattened, with wide confidence intervals reflecting the small number of patients still at risk.

While sex showed no significant association with survival (log-rank $\chi^2_1 = 2.73$, $p = 0.10$, HR for men versus women: $0.53$, $95\%$ CI: $0.24$--$1.14$; $C$-index: $0.57$), age group was associated with it (log-rank $\chi^2_3 = 12.5$, $p = 0.006$). Specifically, the survival curves separated progressively, with lower survival probability in the older groups (Figure~\ref{fig:km-age}). Median survival was not reached in the two youngest groups ($<$55 and 55--65 years) but was reached in the 65--75 and $\geq$75 groups, with lower median and restricted mean survival in the oldest. Relative to the $<$55-year group, although the CIs overlapped, the HR increased with age: HR $= 1.97$ ($95\%$ CI $0.46$--$8.47$) for 55--65 years, HR $= 4.88$ ($95\%$ CI $1.33$--$17.85$) for 65--75 years, and HR $= 7.24$ ($95\%$ CI $1.80$--$29.07$) for $\geq$75 years ($C$-index $= 0.66$). Neither variable showed evidence of violating the PH assumption (Schoenfeld $p = 0.79$ for sex; $p = 0.25$ for age).

After applying Step~1 to the training data of the single split, 7 of the initial 535 radiomic features were selected (Table~\ref{tab:selected_features}). Their FDR-adjusted $p$-values ranged from $0.075$ to $0.132$, and their stability across repeated splits exhibited considerable heterogeneity, with selection frequencies (SF) spanning from $0.06$ to $0.53$ and conditional selection frequencies (CSF) from $0.08$ to $0.74$. These features originated from three tissue compartments: subcutaneous fat ($n = 4$), skeletal muscle ($n = 2$), and intramuscular fat ($n = 1$), with no features selected from either visceral fat or the manual ROI. By radiomic family, NGTDM, Shape, and GLCM were equally represented ($n = 2$ each), followed by a single GLSZM feature (Table~\ref{tab:selected_features}).

\begin{figure*}[!t]
    \centering
    \begin{subfigure}[b]{0.35\textwidth}
        \centering
        \includegraphics[trim=0cm 0cm 0cm 0cm, clip, width=\textwidth]{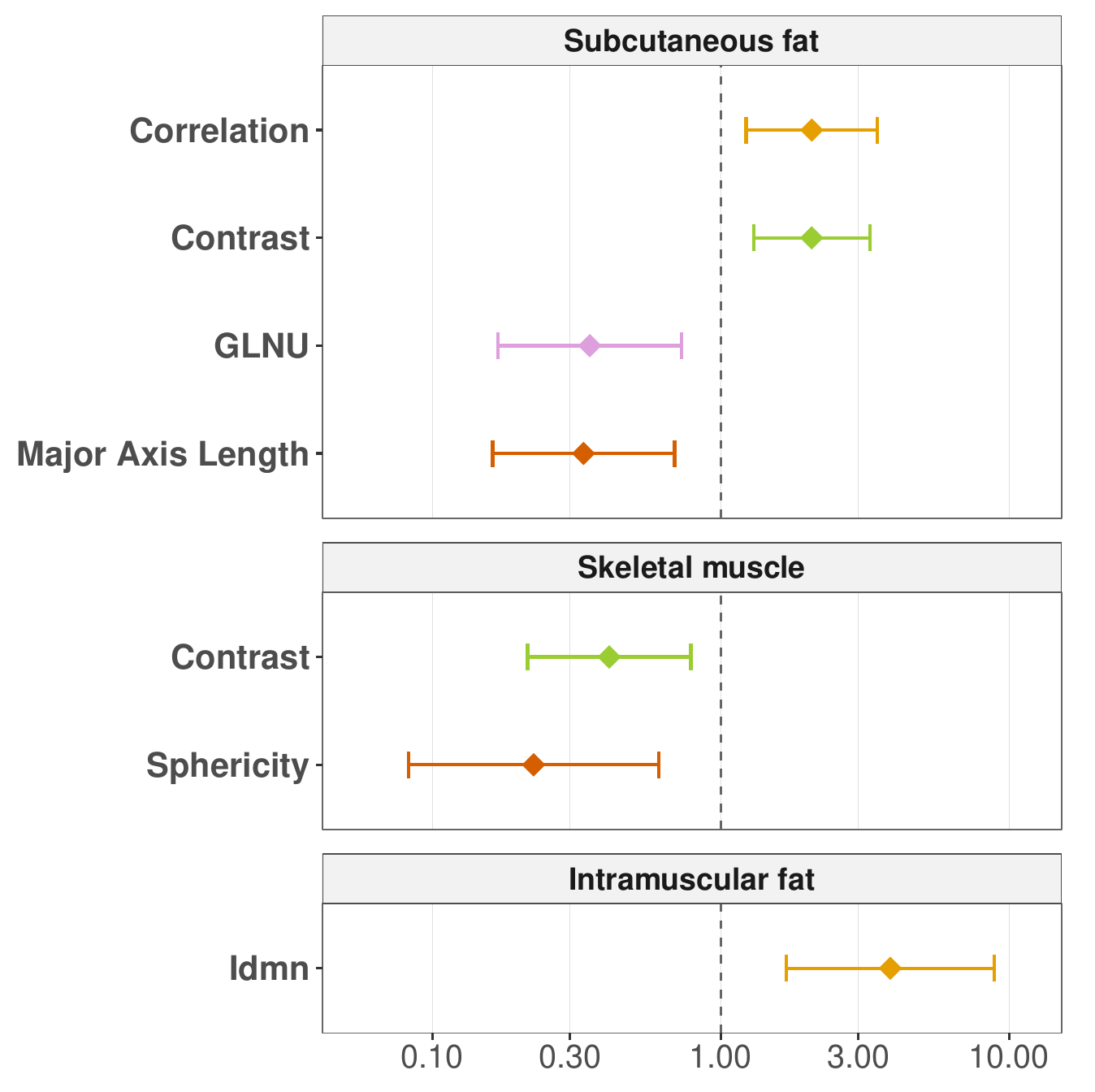}
        \caption{}
        \label{fig:forest}
    \end{subfigure}
    \hfill  
    \begin{subfigure}[b]{0.50\textwidth}
        \centering
        \includegraphics[trim=0cm 0cm 0cm 0cm, clip, width=\textwidth]{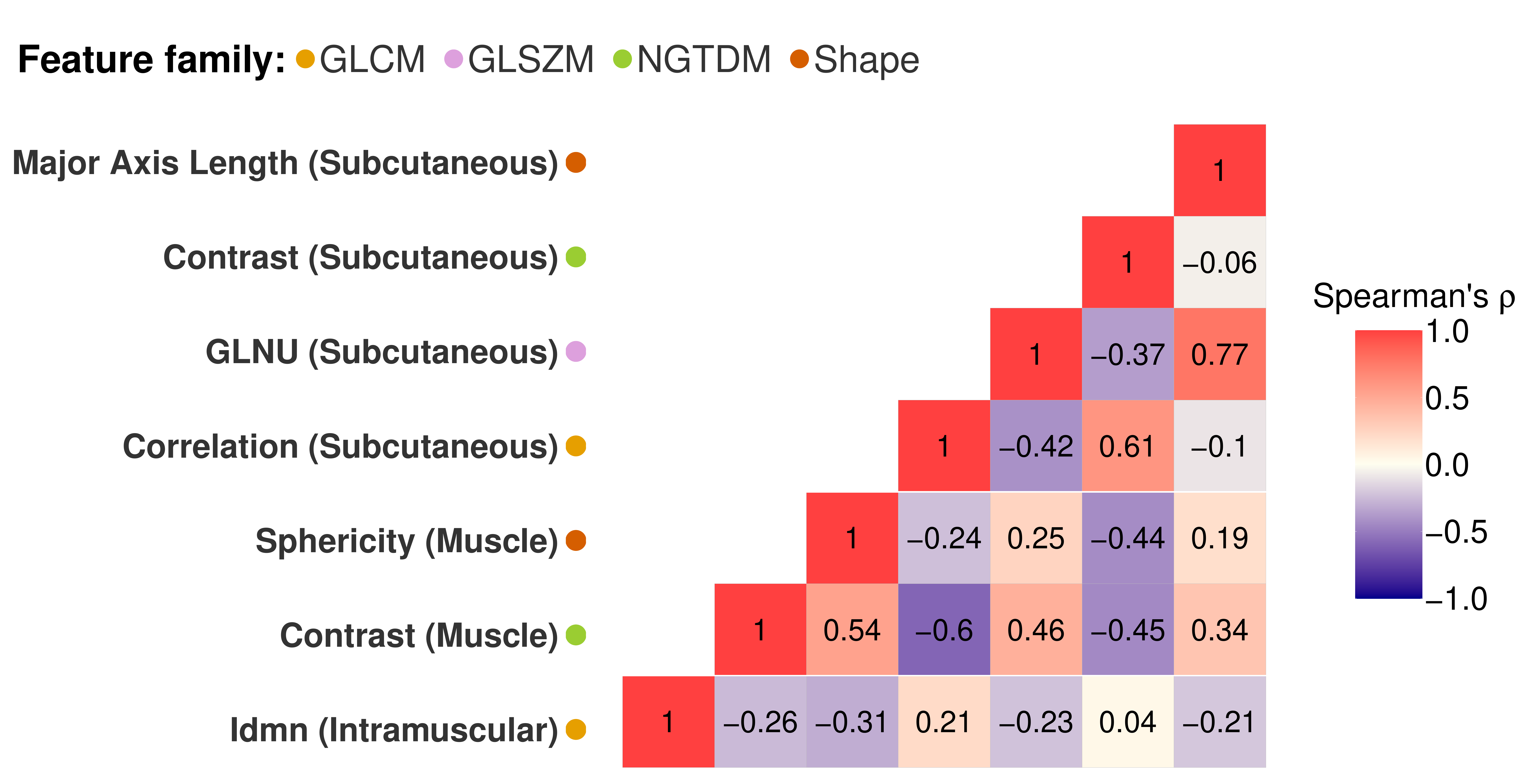}
        \vspace{.25cm}
        \caption{}
        \label{fig:correlation}
    \end{subfigure}
    
    \caption{Univariate Cox analysis and correlation of the selected radiomic features in the training data of the single split. (a) Forest plot of the hazard ratios per SD (HR, Log scale) and $95\%$ confidence intervals (CI), grouped by compartment. (b) Spearman correlation matrix of the 7 selected features, with all pairwise $|\rho|$ below the redundancy filter threshold ($0.80$). \textit{GLNU: Gray-Level Non-Uniformity $\cdot$ Idmn: Inverse Difference Moment Normalized $\cdot$ GLCM: Gray-Level Co-occurrence Matrix $\cdot$ GLSZM: Gray-Level Size Zone Matrix $\cdot$ NGTDM: Neighborhood Gray-Tone Difference Matrix.}}
    \label{fig:figura4}
\end{figure*}

\begin{table*}[!t]
    \centering
    \caption{Radiomic features selected in Step~1 in the training data of the single split (FDR filter: univariate Cox analysis with Benjamini--Hochberg correction, $q \leq 0.20$; redundancy filter: correlation clustering, $|\rho| \geq 0.80$), grouped by compartment. HR: hazard ratio per standard deviation ($\mathrm{HR}<1$ indicates a protective effect); CI: $95\%$ confidence interval; raw $p$ and FDR $p$: unadjusted and Benjamini--Hochberg-adjusted $p$-values from the univariate Cox analysis (Wald test); $C$-index: Harrell's concordance index from the univariate Cox analysis; Schoenfeld $p$: test of the proportional hazards assumption ($p<0.05$ indicates violation); SF and CSF: selection frequency and conditional selection frequency over 100 repetitions of Step~1 (range 0--1).}
    \label{tab:selected_features}
    \footnotesize
    \resizebox{\textwidth}{!}{%
    \begin{tabular}{l l l c c c c c c c}
        \toprule
        \textbf{Feature$^a$} & \textbf{Tissue compartment} & \textbf{Radiomic family$^b$} & \textbf{HR} & \textbf{$95\%$ CI} & \textbf{Raw $p$-value} & \textbf{FDR $p$-value} & \textbf{$C$-index} & \textbf{Schoenfeld $p$} & \textbf{SF/CSF} \\
        \midrule
        Contrast & \multirow[c]{4}{*}{Subcutaneous fat} & NGTDM & 2.07 & 1.30--3.29 & 0.002 & 0.080 & 0.70 & 0.955 & 0.25 / 0.35 \\
        Major Axis Length & & Shape & 0.33 & 0.16--0.69 & 0.003 & 0.088 & 0.64 & 0.021 & 0.25 / 0.35 \\
        GLNU & & GLSZM & 0.35 & 0.17--0.73 & 0.005 & 0.130 & 0.66 & 0.140 & 0.53 / 0.74 \\
        Correlation & & GLCM & 2.07 & 1.23--3.50 & 0.007 & 0.132 & 0.68 & 0.475 & 0.07 / 0.10 \\
        \midrule   
        Sphericity & \multirow[c]{2}{*}{Skeletal muscle} & Shape & 0.22 & 0.08--0.61 & 0.003 & 0.091 & 0.75 & 0.369 & 0.50 / 0.69 \\
        Contrast & & NGTDM & 0.41 & 0.21--0.79 & 0.007 & 0.132 & 0.63 & 0.195 & 0.41 / 0.57 \\
        \midrule
        Idmn & Intramuscular fat & GLCM & 3.87 & 1.69--8.89 & 0.001 & 0.075 & 0.69 & 0.265 & 0.06 / 0.08 \\
        \bottomrule
    \end{tabular}%
    }
    
    \vspace{1ex}
    \raggedright \footnotesize $^a$ GLNU: Gray-Level Non-Uniformity $\cdot$ Idmn: Inverse Difference Moment Normalized.\\
    $^b$ GLCM: Gray-Level Co-occurrence Matrix $\cdot$ GLSZM: Gray-Level Size Zone Matrix $\cdot$ NGTDM: Neighborhood Gray-Tone Difference Matrix.
\end{table*}

Four of the seven features showed a protective effect (HR $<$ 1): Contrast and Sphericity in skeletal muscle, alongside GLNU and Major Axis Length in subcutaneous fat. Conversely, the remaining three features were associated with increased risk (HR $>$ 1): Idmn in intramuscular fat, and Contrast and Correlation in subcutaneous fat (Figure~\ref{fig:forest}). Analysis of the scaled Schoenfeld residuals revealed a violation of the PH assumption exclusively for Major Axis Length in subcutaneous fat (Table~\ref{tab:selected_features}). 
Regarding the selection stability assessment, only two features were selected in at least half of the repetitions (SF $\geq 0.5$): GLNU from subcutaneous fat and Sphericity from skeletal muscle. When considering the CSF, Contrast from skeletal muscle also exceeded this threshold. These statistics are strictly descriptive, as they were derived from an initial feature selection.

Pairwise correlations among the selected features remained below the clustering threshold ($|\rho| < 0.80$), with a maximum correlation of $|\rho| = 0.77$, observed between GLNU and Major Axis Length in subcutaneous fat (Figure~\ref{fig:correlation}).

\subsection{Predictive performance estimates}

\begin{figure*}[!t]
    \centering
    \begin{subfigure}[t]{0.44\textwidth}
        \centering
        \includegraphics[trim=0cm -1.5cm 0cm 0cm, clip, width=\textwidth]{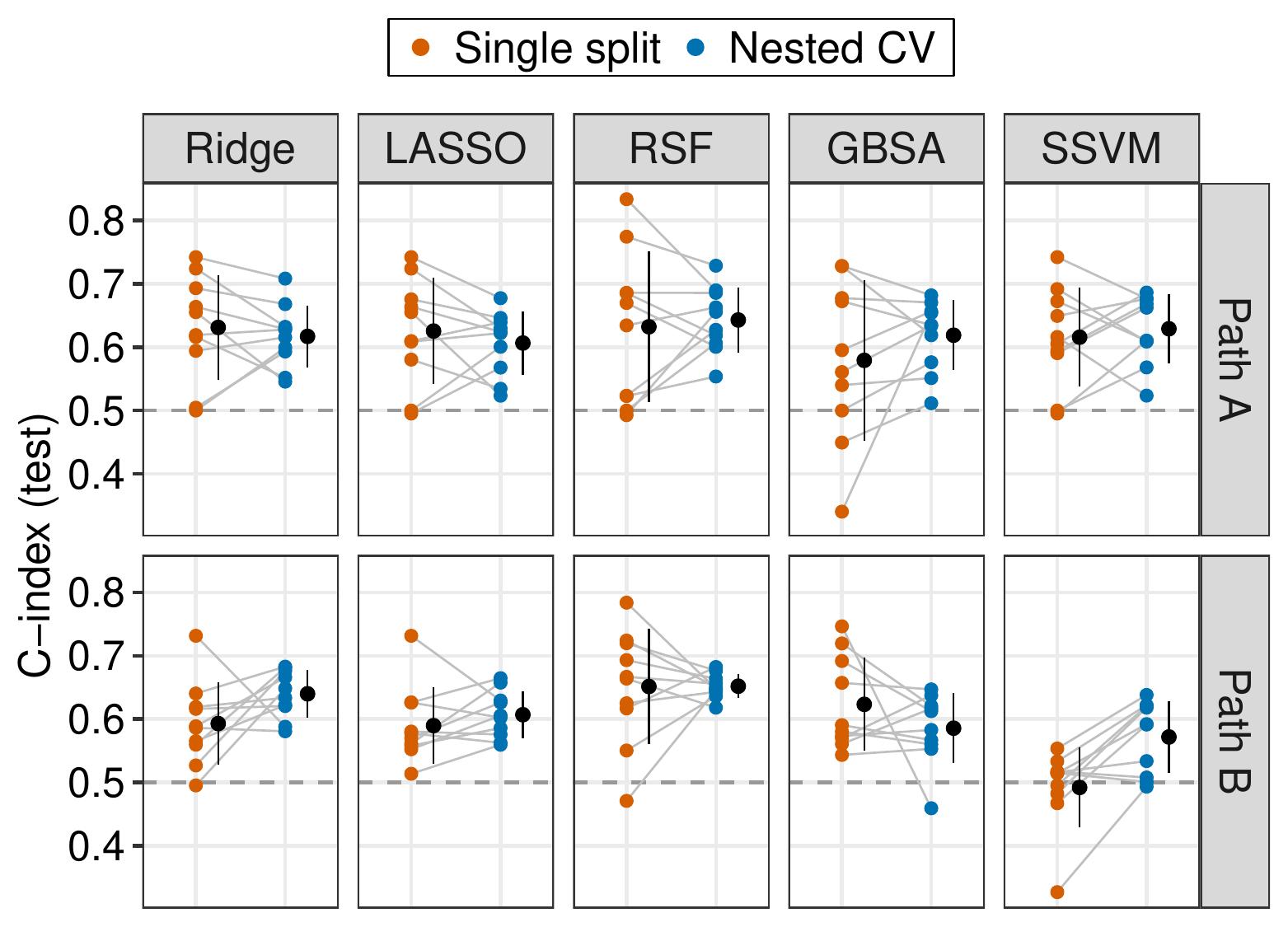}
        \caption{}
        \label{fig:single_variability}
    \end{subfigure}
    \hspace{0.05\textwidth}%
    \begin{subfigure}[t]{0.41\textwidth}
        \centering
        \includegraphics[trim=0cm 0cm 0cm 0cm, clip, width=\linewidth]{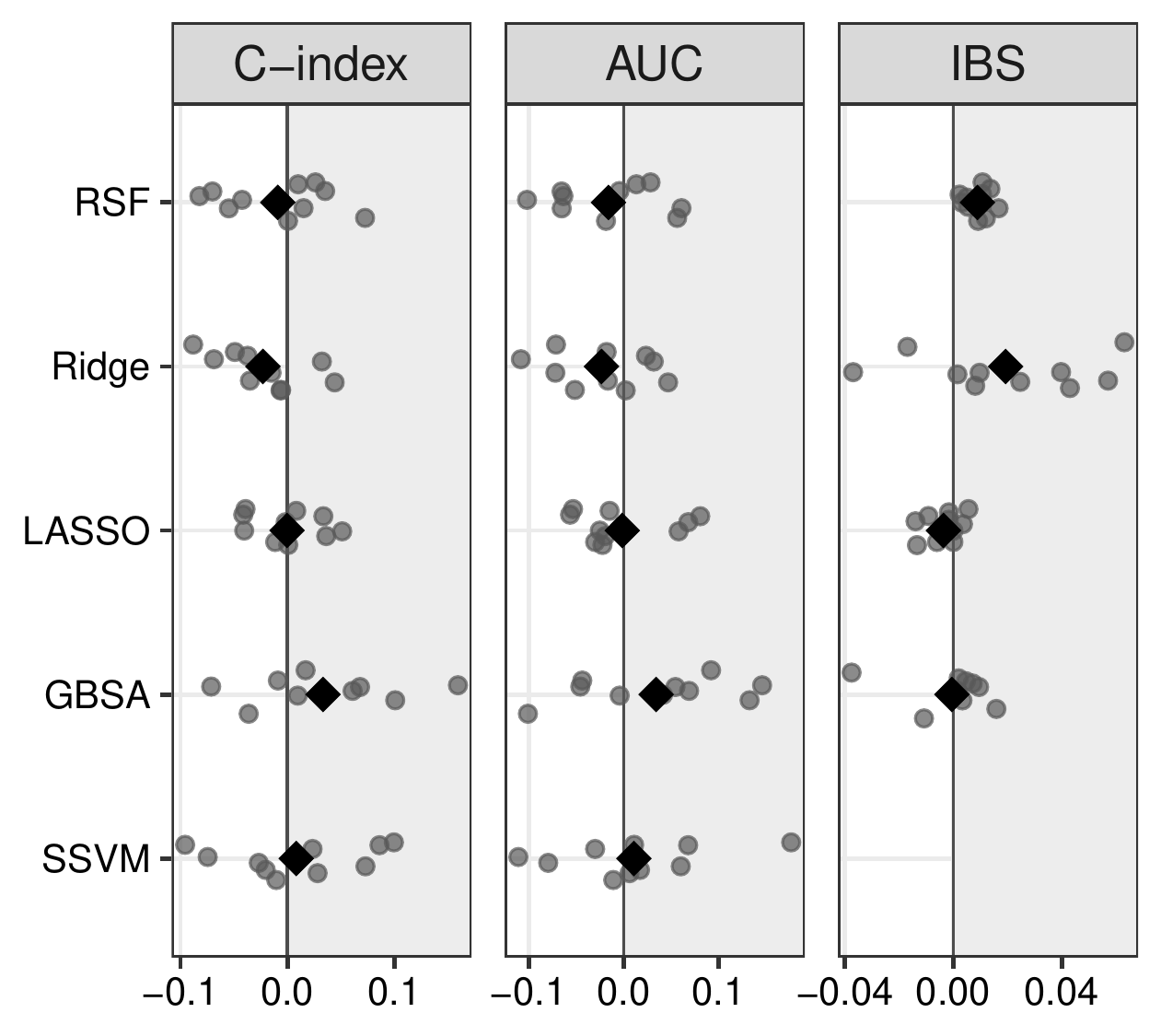}
        \caption{}
        \label{fig:path_a_adv}
    \end{subfigure}
    \caption{Performance of the survival models under the two validation schemes (a) and comparison between Path~A (prior feature selection) and Path~B (full set of features with embedded regularization) (b). (a)~Harrell's concordance index ($C$-index) obtained with the single split (orange) and with repeated nested cross-validation (CV) (blue), for each model--path combination. Each gray line joins, for one seed, the single-split estimate and the nested CV estimate (mean across the five outer folds); black points and bars show the mean $\pm$ standard deviation (SD) between seeds, and the dashed line marks chance level ($C$-index $=0.5$). Numerical values are given in Table~\ref{tab:single_vs_nested_full}. (b)~Per-seed paired difference between Path~A and Path~B for the $C$-index, the time-dependent area under the receiver operating characteristic curve (AUC) and the integrated Brier score (IBS). Gray points are individual seeds ($n=10$) and black diamonds the mean across seeds ($\Delta_{A-B}$). Positive values (shaded area) indicate better performance of Path~A (sign inverted for the IBS, which is undefined for SSVM). Horizontal axes are scaled independently for each metric. Numerical values are given in Table~\ref{tab:dual_path}.}
    \label{fig:path_adv_variability}
\end{figure*}

Under repeated nested CV with the radiomic features, the mean $C$-index of the model--path combinations ranged from $0.572$ to $0.652$ and the mean time-dependent AUC from $0.577$ to $0.663$, both above the chance level of $0.5$, while the mean IBS stayed below the reference value of $0.25$ in all combinations for which it is defined (Table~\ref{tab:single_vs_nested_full}). Single-split estimates, however, deviated substantially from these values: across seeds, the mean absolute difference between the two schemes ($|\Delta|$) was $0.069$ for the $C$-index, $0.100$ for the AUC and $0.028$ for the IBS, of the same order as the range of the model means under nested CV. These differences showed no overall direction, as the mean difference ($\Delta$) took both signs across models and was close to zero on average. The largest discrepancies occurred for SSVM in Path~B, where the single-split estimates fell below the nested CV estimates by more than the SD between seeds of both schemes (Table~\ref{tab:single_vs_nested_full}).
Rather than in direction, the two schemes differed in the dispersion of their estimates (Figure~\ref{fig:single_variability}). The SD between seeds of the single-split estimates, averaged across models and paths, was $0.084$ for the $C$-index, $0.101$ for the AUC and $0.038$ for the IBS, comparable to the range of the model means. Nested CV reduced this SD by a median of $41\%$, $28\%$ and $60\%$, respectively, although the remaining dispersion was still of the same order as the differences between models.

\begin{table*}[!t]
\centering
\caption{Performance of each model--path combination over ten seeds, estimated with repeated nested cross-validation (CV) and the single split on the radiomic features, and with repeated nested CV only on the Gaussian noise features. Performance was measured with Harrell's concordance index ($C$-index), the time-dependent area under the receiver operating characteristic curve (AUC) and the integrated Brier score (IBS). The nested CV and single split columns show the mean $\pm$ standard deviation (SD) between seeds, accounting for folds with an empty selection\textsuperscript{a}. Upward arrows indicate metrics where higher values represent better performance, whereas the downward arrow indicates the opposite.}
\label{tab:single_vs_nested_full}
\footnotesize
\setlength{\tabcolsep}{4pt}
\begin{tabular}{lll cccccc}
\toprule
 & & & \multicolumn{4}{c}{Radiomic features} & \multicolumn{2}{c}{Gaussian noise features} \\
\cmidrule(lr){4-7} \cmidrule(lr){8-9}
Metric & Path & Model & Nested CV & Single split & $\Delta$\textsuperscript{b, d} & $|\Delta|$\textsuperscript{c, d} & Nested CV & Margin\textsuperscript{e} \\
\midrule
\multirow{10}{*}{\rotatebox[origin=c]{90}{\textbf{$C$-index $\uparrow$}}}
 & \multirow{5}{*}{A} & Cox Ridge & $0.617 \pm 0.049$ & $0.631 \pm 0.083$ & $+0.014$ & $0.058$ & $0.405 \pm 0.048$ & $0.231$ \\
 & & Cox LASSO & $0.606 \pm 0.050$ & $0.625 \pm 0.084$ & $+0.019$ & $0.062$ & $0.413 \pm 0.051$ & $0.210$ \\
 & & RSF & $0.643 \pm 0.051$ & $0.632 \pm 0.119$ & $-0.011$ & $0.076$ & $0.423 \pm 0.073$ & $0.245$ \\
 & & GBSA & $0.619 \pm 0.055$ & $0.579 \pm 0.127$ & $-0.040$ & $0.080$ & $0.412 \pm 0.037$ & $0.226$ \\
 & & SSVM & $0.629 \pm 0.054$ & $0.616 \pm 0.078$ & $-0.013$ & $0.074$ & $0.399 \pm 0.037$ & $0.250$ \\
\cmidrule(lr){2-9}
 & \multirow{5}{*}{B} & Cox Ridge & $0.640 \pm 0.037$ & $0.593 \pm 0.065$ & $-0.047$ & $0.077$ & $0.553 \pm 0.093$ & $0.087$ \\
 & & Cox LASSO & $0.607 \pm 0.037$ & $0.590 \pm 0.060$ & $-0.017$ & $0.046$ & $0.492 \pm 0.082$ & $0.115$ \\
 & & RSF & $0.652 \pm 0.019$ & $0.651 \pm 0.091$ & $0.000$ & $0.068$ & $0.525 \pm 0.039$ & $0.127$ \\
 & & GBSA & $0.586 \pm 0.055$ & $0.623 \pm 0.074$ & $+0.038$ & $0.066$ & $0.473 \pm 0.065$ & $0.113$ \\
 & & SSVM & $0.572 \pm 0.057$ & $0.492 \pm 0.063$ & $-0.080$ & $0.084$ & $0.545 \pm 0.084$ & $0.027$ \\
\midrule
\multirow{10}{*}{\rotatebox[origin=c]{90}{\textbf{AUC $\uparrow$}}}
 & \multirow{5}{*}{A} & Cox Ridge & $0.625 \pm 0.073$ & $0.640 \pm 0.095$ & $+0.028$ & $0.093$ & $0.388 \pm 0.056$ & $0.260$ \\
 & & Cox LASSO & $0.623 \pm 0.073$ & $0.634 \pm 0.097$ & $+0.023$ & $0.093$ & $0.390 \pm 0.056$ & $0.254$ \\
 & & RSF & $0.647 \pm 0.073$ & $0.651 \pm 0.137$ & $+0.019$ & $0.116$ & $0.411 \pm 0.088$ & $0.266$ \\
 & & GBSA & $0.614 \pm 0.082$ & $0.616 \pm 0.150$ & $+0.013$ & $0.120$ & $0.417 \pm 0.061$ & $0.218$ \\
 & & SSVM & $0.624 \pm 0.086$ & $0.642 \pm 0.089$ & $+0.027$ & $0.093$ & $0.391 \pm 0.070$ & $0.251$ \\
\cmidrule(lr){2-9}
 & \multirow{5}{*}{B} & Cox Ridge & $0.649 \pm 0.058$ & $0.598 \pm 0.092$ & $-0.044$ & $0.100$ & $0.576 \pm 0.098$ & $0.073$ \\
 & & Cox LASSO & $0.625 \pm 0.074$ & $0.633 \pm 0.082$ & $+0.014$ & $0.097$ & $0.498 \pm 0.102$ & $0.127$ \\
 & & RSF & $0.663 \pm 0.047$ & $0.678 \pm 0.108$ & $+0.021$ & $0.105$ & $0.530 \pm 0.043$ & $0.133$ \\
 & & GBSA & $0.580 \pm 0.056$ & $0.635 \pm 0.082$ & $+0.064$ & $0.069$ & $0.478 \pm 0.081$ & $0.102$ \\
 & & SSVM & $0.577 \pm 0.070$ & $0.473 \pm 0.077$ & $-0.111$ & $0.111$ & $0.541 \pm 0.100$ & $0.036$ \\
\midrule
\multirow{10}{*}{\rotatebox[origin=c]{90}{\textbf{IBS $\downarrow$}\textsuperscript{f}}}
 & \multirow{5}{*}{A} & Cox Ridge & $0.189 \pm 0.018$ & $0.186 \pm 0.034$ & $+0.003$ & $0.029$ & $0.249 \pm 0.036$ & $0.059$ \\
 & & Cox LASSO & $0.192 \pm 0.014$ & $0.187 \pm 0.020$ & $+0.006$ & $0.022$ & $0.227 \pm 0.038$ & $0.034$ \\
 & & RSF & $0.175 \pm 0.012$ & $0.178 \pm 0.027$ & $-0.003$ & $0.021$ & $0.237 \pm 0.038$ & $0.061$ \\
 & & GBSA & $0.192 \pm 0.024$ & $0.196 \pm 0.045$ & $-0.004$ & $0.029$ & $0.230 \pm 0.022$ & $0.038$ \\
 & & SSVM & -- & -- & -- & -- & -- & -- \\
\cmidrule(lr){2-9}
 & \multirow{5}{*}{B} & Cox Ridge & $0.208 \pm 0.027$ & $0.241 \pm 0.092$ & $-0.033$ & $0.060$ & $0.224 \pm 0.034$ & $0.016$ \\
 & & Cox LASSO & $0.189 \pm 0.008$ & $0.186 \pm 0.022$ & $+0.003$ & $0.016$ & $0.200 \pm 0.009$ & $0.011$ \\
 & & RSF & $0.181 \pm 0.008$ & $0.175 \pm 0.022$ & $+0.006$ & $0.017$ & $0.191 \pm 0.008$ & $0.010$ \\
 & & GBSA & $0.191 \pm 0.010$ & $0.198 \pm 0.043$ & $-0.007$ & $0.026$ & $0.203 \pm 0.011$ & $0.012$ \\
 & & SSVM & -- & -- & -- & -- & -- & -- \\
\bottomrule
\end{tabular}
\vspace{3pt}
\begin{minipage}{\textwidth}
\footnotesize
\textsuperscript{a}\, For Path~A, folds in which Step~1 selected no features were assigned a $C$-index and AUC of $0.5$ and excluded from the IBS. With the radiomic features, this affected $8$ of $50$ folds. Under noise, Path~A estimates are instead restricted to the folds with a non-empty selection ($9$ of $50$ folds), so each seed value is the mean over its non-empty folds.\\
\textsuperscript{b}\,Mean difference between schemes, $\Delta=\frac{1}{n}\sum_{s}\left(S_s-\bar{N}_s\right)$, where $S_s$ is the single-split estimate for seed $s$ and $\bar{N}_s$ the nested CV estimate (mean across the five outer folds) for that seed. For the IBS, the sign is inverted ($\Delta=\frac{1}{n}\sum_{s}\left(\bar{N}_s-S_s\right)$).\\
\textsuperscript{c}\,Mean absolute difference between schemes, $|\Delta|=\frac{1}{n}\sum_{s}\left|S_s-\bar{N}_s\right|$.\\
\textsuperscript{d}\,$\Delta$ and $|\Delta|$ were computed only over the seeds in which both estimates were defined (all ten seeds, except for the AUC and the IBS in Path~A, where nine were available). Consequently, $\Delta$ can differ from the difference between the two displayed columns.\\
\textsuperscript{e}\,Margin over noise, $N_{\mathrm{rad}}-N_{\mathrm{noise}}$, where $N_{\mathrm{rad}}$ and $N_{\mathrm{noise}}$ are the nested CV means across seeds with the radiomic and Gaussian noise features, respectively, computed from unrounded values. Positive values indicate better performance with the radiomic features, so the sign was inverted for the IBS ($N_{\mathrm{noise}}-N_{\mathrm{rad}}$).
For Path~A, both estimates are restricted to the folds with a non-empty selection (see note~a); Path~B has no empty folds. Consequently, for the $C$-index and AUC, the Path~A margin can differ from the difference between the displayed Nested CV columns (for instance, Cox Ridge, $C$-index: $0.636-0.405=0.231$, versus $0.617-0.405=0.212$ from the displayed columns).\\
\textsuperscript{f}\,IBS is undefined for SSVM, which does not provide survival functions.\\
\end{minipage}
\end{table*}

Turning to the dual-path comparison, no metric showed a consistent difference between paths: the mean paired difference ($\Delta_{A-B}$) was close to zero when averaged across models, and in 13 of the 14 model--metric combinations it was smaller than its SD between seeds (Table~\ref{tab:dual_path}). The IBS was the only metric showing a pattern, favoring Path~A (lower IBS) in all ten seeds for RSF and in eight of ten for Cox Ridge (Figure~\ref{fig:path_a_adv}), although the mean difference exceeded its SD between seeds only for RSF. For these two models, in contrast, the $C$-index and AUC were on average higher in Path~B, with differences smaller than the SD between seeds.

\begin{table*}[!t]
\centering
\caption{Dual-path comparison under repeated nested cross-validation (CV) over ten seeds, in which Path~A fits the models on the selected set of features (prior feature selection) and Path~B on the full set of features (embedded regularization). Performance was measured with Harrell's concordance index ($C$-index), the time-dependent area under the receiver operating characteristic curve (AUC) and the integrated Brier score (IBS). $\Delta_{A-B}$\textsuperscript{a} is the mean paired difference between paths ($\pm$ standard deviation, SD, between seeds), and $n_A$\textsuperscript{b} is the number of seeds in which Path~A performed better. Positive $\Delta_{A-B}$ values indicate better performance of Path~A for all three metrics. Upward arrows indicate metrics for which higher values are better, whereas the downward arrow indicates the opposite. The mean estimate of each path is reported in Table~\ref{tab:single_vs_nested_full}.}
\label{tab:dual_path}
\footnotesize
\setlength{\tabcolsep}{6pt}
\begin{tabular}{l cc cc cc}
\toprule
 & \multicolumn{2}{c}{\textbf{$C$-index}\,$\uparrow$} & \multicolumn{2}{c}{\textbf{AUC}\,$\uparrow$} & \multicolumn{2}{c}{\textbf{IBS}\textsuperscript{c}\,$\downarrow$} \\
\cmidrule(lr){2-3} \cmidrule(lr){4-5} \cmidrule(lr){6-7}
Model & $\Delta_{A-B}$ & $n_A$ & $\Delta_{A-B}$ & $n_A$ & $\Delta_{A-B}$ & $n_A$ \\
\midrule
Cox Ridge & $-0.023 \pm 0.042$ & 2/10 & $-0.023 \pm 0.051$ & 4/10 & $+0.019 \pm 0.032$ & 8/10 \\
Cox LASSO & $0.000 \pm 0.034$ & 5/10 & $-0.002 \pm 0.051$ & 3/10 & $-0.004 \pm 0.007$ & 3/10 \\
RSF       & $-0.009 \pm 0.051$ & 6/10 & $-0.016 \pm 0.056$ & 4/10 & $+0.009 \pm 0.005$ & 10/10 \\
GBSA      & $+0.033 \pm 0.067$ & 7/10 & $+0.034 \pm 0.081$ & 6/10 & $-0.001 \pm 0.015$ & 7/10 \\
SSVM      & $+0.057 \pm 0.082$ & 8/10 & $+0.047 \pm 0.100$ & 8/10 & -- & -- \\
\bottomrule
\end{tabular}
\vspace{4pt}
\begin{minipage}{\textwidth}
\footnotesize
\textsuperscript{a}\,Mean paired difference between paths ($\Delta_{A-B}$). For each seed $s$, Path~A and Path~B were compared in every outer fold, and the differences were averaged across the folds of that seed ($\bar{d}_s$). $\Delta_{A-B}$ is the mean of $\bar{d}_s$ across the $n$ seeds, $\Delta_{A-B}=\frac{1}{n}\sum_{s}\bar{d}_s$.
Positive values indicate better performance of Path~A, so the sign was inverted for the IBS ($B_{s,f}-A_{s,f}$).
For the IBS, folds in which Step~1 selected no features are excluded from $\bar{d}_s$, so $\Delta_{A-B}$ and $n_A$ need not equal the difference between the path means in Table~\ref{tab:single_vs_nested_full}. All differences were computed from unrounded values.\\
\textsuperscript{b}\,Number of seeds in which Path~A performed better ($\bar{d}_s>0$).\\
\textsuperscript{c}\,IBS is undefined for SSVM, which does not provide survival functions.
\end{minipage}
\end{table*}

Regarding the negative control, Step~1 selected at least one feature in $18\%$ of the outer folds with Gaussian noise features and in $84\%$ with the radiomic features. A similar pattern was observed in the single split, where Step~1 selected no features in $9$ of the $10$ seeds with Gaussian noise features and in $1$ of $10$ with the radiomic features. In Path~A with the radiomic features, the outer folds with an empty selection, which were assigned chance-level performance, lowered the nested CV $C$-index and AUC by $0.02$--$0.03$ relative to the estimates over the non-empty folds (Table~\ref{tab:single_vs_nested_full}).

\noindent
\begin{minipage}{\columnwidth}
    \captionsetup{type=figure, font={small,sf}, labelfont=bf, labelsep=colon}
    \centering
    \includegraphics[trim=-1.2cm 0cm 0cm 0cm, clip, width=\linewidth]{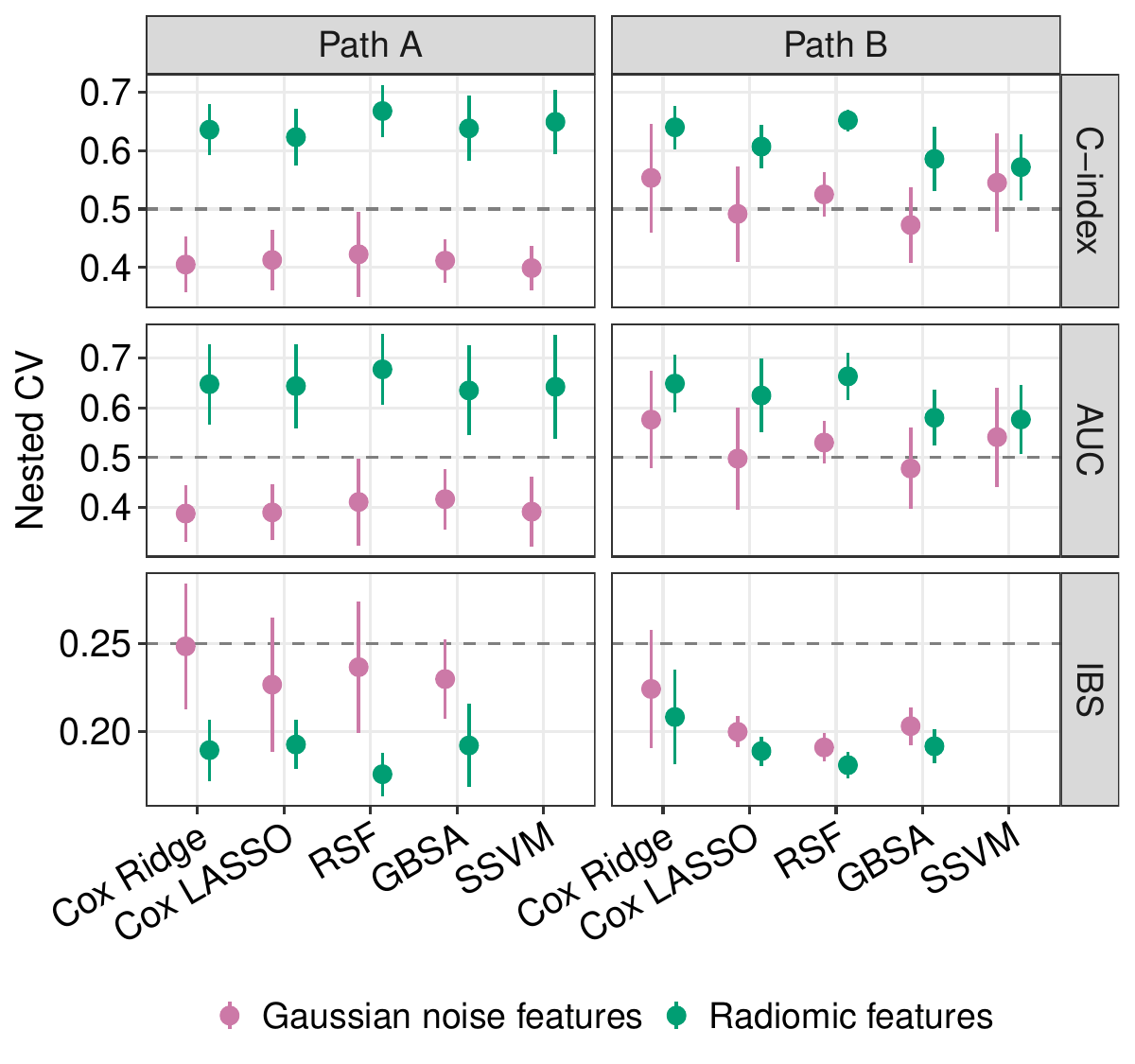}
    \caption{Negative control: performance of the survival models under repeated nested cross-validation (CV) with the radiomic features and with Gaussian noise features, for each model in Path~A (prior feature selection, left) and Path~B (full set of features with embedded regularization, right). Points and bars show the mean $\pm$ standard deviation (SD) between seeds of Harrell's concordance index ($C$-index), the time-dependent area under the receiver operating characteristic curve (AUC) and the integrated Brier score (IBS, undefined for SSVM). Dashed lines mark chance level ($0.5$ for the $C$-index and AUC) and the reference value ($0.25$ for the IBS). For Path~A, estimates are restricted to the folds in which Step~1 selected at least one feature. Numerical values and fold counts are given in Table~\ref{tab:single_vs_nested_full} (note~a).}
    \label{fig:noisy_vs_real}
\end{minipage}
\vspace{0.5em}

In terms of performance with Gaussian noise features, neither path exceeded chance level beyond the variability between seeds (Figure~\ref{fig:noisy_vs_real}). In Path~B, the $C$-index ranged from $0.473$ to $0.553$, within one SD between seeds of $0.5$ in every model. In Path~A, restricted to the folds in which Step~1 selected at least one feature (nine folds and six seeds), the $C$-index remained below chance level ($0.399$--$0.423$). The AUC led to the same conclusions in both paths. The margin of the radiomic features over noise was larger in Path~A ($C$-index $0.21$--$0.25$) than in Path~B ($0.03$--$0.13$), and the IBS showed the same pattern: under noise it was higher (worse) than with the radiomic features by at least $0.034$ in Path~A (over the folds in which a model was fitted) but by at most $0.016$ in Path~B (Table~\ref{tab:single_vs_nested_full}, Figure~\ref{fig:noisy_vs_real}).

\FloatBarrier
\section{Discussion}

\subsection{Biological relevance of selected features}

Three features were selected considerably more often than the rest (CSF $\geq 0.5$), all of them associated with lower risk: GLNU in subcutaneous fat, and Sphericity and Contrast in skeletal muscle (Table~\ref{tab:selected_features}). In the latter compartment, the protective effect of Sphericity may reflect a more compact shape, whereas that of Contrast, which captures the spatial rate of intensity change, may reflect a better-preserved texture.
In subcutaneous fat, GLNU tends to increase with the number of size zones and hence with the amount of tissue, which suggests that a greater subcutaneous tissue extent may relate to a better prognosis.
Contrast in subcutaneous fat, which was selected less often than the three features above (CSF of $0.35$), was associated with higher risk. This may indicate that a greater spatial variation of intensity in this compartment reflects a heterogeneity linked to adverse outcome, unlike in skeletal muscle. The meaning of this feature may therefore differ between compartments. Given its lower selection frequency, however, this association is less reliable and chance cannot be excluded.
Major Axis Length in subcutaneous fat was selected with a similar frequency, but it violated the PH assumption (Schoenfeld $p = 0.021$), so its hazard ratio was not interpreted. The remaining two features, Correlation in subcutaneous fat and Idmn in intramuscular fat, were selected too rarely to be interpreted (Table~\ref{tab:selected_features}).

Taken together, the features selected most often point to subcutaneous tissue extent, muscle compactness and muscle texture as tentative correlates of a more favorable prognosis. These findings are exploratory, however, given the modest FDR-adjusted $p$-values of the selected features (Table~\ref{tab:selected_features}) and their heterogeneous selection frequencies, and should be regarded as hypothesis-generating.

\subsection{Feature selection and validation protocol}

The difference between the single-split and nested CV estimates showed no systematic direction, but it was not negligible: the mean absolute difference between schemes ($|\Delta|$) was $0.07$ for the $C$-index (Table~\ref{tab:single_vs_nested_full}).

The SD between seeds of the single-split estimates was comparable to the range of the model means under nested CV for the three metrics (Table~\ref{tab:single_vs_nested_full}). A model comparison based on a single split can therefore reflect the split drawn rather than genuine differences between models. Repeated nested CV reduced this variability (median SD reduction of $41\%$, $28\%$, and $60\%$ for the $C$-index, AUC, and IBS, respectively) but did not eliminate it: the remaining SD between seeds ($0.02$--$0.06$ for the $C$-index, Table~\ref{tab:single_vs_nested_full}) was still of the same order as the differences between models, in line with the fact that no model could be clearly distinguished from the others.

The dual-path comparison served as a sensitivity analysis of the dimensionality-reduction strategy. With the radiomic features, the mean nested CV $C$-index exceeded chance level by $0.07$--$0.15$ across models and paths, a modest signal of similar magnitude in both paths. Neither path was consistently better: differences between them were small relative to their dispersion and their direction varied between models. We therefore found no evidence that performance depended on the dimensionality-reduction strategy, and prior feature selection offered no consistent predictive benefit.

The negative control showed that, in neither path, performance with Gaussian noise features clearly exceeded chance level. Estimates stayed close to $0.5$ in Path~B ($C$-index $0.47$--$0.55$, within one SD between seeds in every model) and fell below it in Path~A ($C$-index $0.40$--$0.42$). The margin of the radiomic features over noise was therefore small in Path~B ($C$-index $0.03$--$0.13$, AUC $0.04$--$0.13$ across models) and larger in Path~A ($C$-index $0.21$--$0.25$, AUC $0.22$--$0.27$), reflecting noise estimates below $0.5$ (Figure~\ref{fig:noisy_vs_real}). These estimates are available only for the nine outer folds with a non-empty selection and are consistent with the selection of features that do not generalize, so the larger margin should not be interpreted as an advantage of Path~A.

The negative control also showed that, with Gaussian noise features, Step~1 selected at least one feature in only $18\%$ of the outer folds, in line with the upper bound of $q = 0.20$ expected for the Benjamini--Hochberg procedure under a global null hypothesis. With the radiomic features, this proportion was $66$ percentage points higher, which suggests that, in this cohort, at least some features are associated with survival in the training data, beyond what would be expected by chance after FDR correction. The cost of Step~1 was modest: the $16\%$ of outer folds without selected features lowered the nested CV $C$-index and AUC of Path~A by $0.02$--$0.03$. Taken together, the above-chance performance with the radiomic features, its absence under noise and the marked difference in the selection rate point to a modest prognostic signal in this cohort, although its magnitude remains uncertain.

Finally, the metrics captured different aspects of performance \cite{hoessly2026how}. With the radiomic features, the IBS showed a pattern that was not mirrored by the discrimination metrics: for Cox Ridge and RSF, the $C$-index and AUC were on average higher in Path~B, whereas the IBS favored Path~A (lower IBS) in eight of ten seeds for Cox Ridge and in all ten for RSF, although the mean difference exceeded its SD between seeds only for RSF. Given the small magnitude of these differences, we regard this as a tentative signal rather than as evidence of the superiority of Path~A.

\subsection{Limitations and future work}
\label{sec:limitations}

Some limitations should be acknowledged when interpreting these findings.

First, patient recruitment was restricted to a single center using a uniform imaging protocol. While this homogeneity helps reduce technical confounding, it also limits the generalizability of our results. Although the repeated nested CV reduces the dependence of the estimates on a single split, it cannot fully replace external validation, the gold standard for assessing generalizability \cite{klontzas2024radiomics}. Such external validation was not performed in this study, as, to the best of our knowledge, there is no publicly available dataset combining CT-based L3 body composition with GEP-NET survival outcomes.

Second, because the results derive from a single, relatively small cohort, the model-specific differences between Path~A and Path~B should not be extrapolated to other datasets. The main contribution of this study is a protocol that quantifies how the validation scheme and the dimensionality-reduction strategy affect the performance estimates, using a negative control as reference. This is particularly relevant in small cohorts.

Third, the radiomic analysis was based on features extracted from a single baseline CT scan. Incorporating features from follow-up imaging as time-varying covariates could capture longitudinal changes in body composition, potentially improving predictive performance beyond what a single baseline assessment can offer \cite{abbas2023delta}. However, relying on baseline data can be viewed as more clinically relevant, given that it better reflects the real-world decision-making environment prior to the manifestation of the outcome.

Fourth, the negative control of Path~A rests on only nine outer folds (six seeds), so estimates below chance should not be over-interpreted. In addition, assigning chance-level discrimination ($C$-index and AUC of $0.5$) to the folds with an empty selection is a convention of the protocol that shifts the mean estimates of Path~A towards $0.5$ (lowering them by $0.02$--$0.03$ for the radiomic features), and other conventions could yield different values.

Finally, Step~1 selected no features from the manual ROI. This should not be taken as evidence that manual segmentation is less informative than automated whole-muscle segmentation, since the ROI was small, covered a different region, and its shape features may mainly reflect the placement of the segmentation sphere. Given the low statistical power in this compartment, this observation is at most hypothesis-generating.

Building upon these limitations, several natural extensions emerge for future research. Clinically, the skeletal muscle index (SMI; muscle area normalized by height squared) could be derived from the existing L3 segmentations. This would allow relating the radiomic features to low muscle mass, a CT-based proxy of sarcopenia, and determine whether they offer prognostic value beyond this conventional metric.
Methodologically, the comparisons rely on descriptive differences across ten seeds rather than on formal testing, since all seeds resample the same cohort and standard paired tests assume independence. Tests corrected for dependent resampling, such as the Nadeau and Bengio corrected $t$-test \cite{nadeau2003inference}, are a natural extension. A complementary negative control based on permuting the survival outcomes, which preserves the correlation structure of the radiomic features, would also help to confirm the behavior of both paths under the null.

\section{Conclusion}

This study proposed a four-step validation protocol for radiomic survival analysis in small cohorts, in which each step targets a distinct methodological risk: Step~1, FDR-corrected feature selection (false discoveries); Step~2, a dual-path comparison (sensitivity to the dimensionality-reduction strategy); Step~3, repeated nested CV (optimistic or highly variable performance estimates); and Step~4, a negative control on Gaussian noise features (artifacts of the analytical pipeline). The protocol was applied to 78 GEP-NET patients, using 535 CT-based radiomic features extracted from five compartments at the L3 level, to evaluate their prognostic value across five survival models.

In this cohort, each step of the protocol provided useful information for interpreting the performance estimates. Step~1 behaved as designed: it selected at least one feature in only $18\%$ of the outer folds with Gaussian noise features, consistent with its nominal level ($q \leq 0.20$), versus $84\%$ with the radiomic features. The dual-path comparison, a sensitivity analysis of the dimensionality-reduction strategy, showed that neither path was consistently better, so the estimates were largely insensitive to this choice. Compared with a single split, repeated nested CV did not shift the estimates in a consistent direction, but it reduced their variability between seeds (median reduction in SD of $41\%$ for the $C$-index, $28\%$ for the AUC and $60\%$ for the IBS), although the remaining variability was of the same order as the differences between models. Finally, the negative control showed that, whereas the mean nested CV $C$-index with the radiomic features exceeded chance level by $0.07$--$0.15$ across models and paths, performance with Gaussian noise features did not clearly exceed it ($C$-index $0.40$--$0.42$ in Path~A, over nine folds, and $0.47$--$0.55$ in Path~B), so the pipeline itself is unlikely to generate apparent performance from noise.

Reporting several summary measures was also worthwhile \cite{hoessly2026how, dubray2025internal}, as discrimination and overall prediction error did not always agree: for RSF and Cox Ridge, the IBS was lower in Path~A, whereas the $C$-index and AUC were on average higher in Path~B.

Regarding the prognostic analysis, age at diagnosis was associated with overall survival in this cohort. Among the radiomic features, those selected most often pointed, tentatively, to a greater subcutaneous fat extent, a more compact skeletal muscle and preserved muscle texture as correlates of a more favorable prognosis. Given the sample size and the heterogeneous selection frequencies, these interpretations should be regarded as hypothesis-generating.

Taken together, these results suggest that L3 body composition radiomics may carry a modest prognostic signal in this GEP-NET cohort, although the uncertainty quantified by the protocol precludes firm conclusions about any specific model. The protocol is not specific to this application and may be applicable to other radiomic survival studies in small cohorts, where high dimensionality and a limited number of events pose similar methodological risks. Its transferability, however, remains to be tested.

\section*{Funding statement}
This work was supported by the Agencia Estatal de Investigación of Spain under Grant PID2025-171806OB-I00, by the Regional Government of Madrid of Spain under Grant TEC-2024/COM-322, and by the FPI-UAM grant.

\section*{Ethics statement}
This study was approved by the Research Ethics Committee (CEIm) of Hospital Universitario de La Princesa, Madrid, Spain (registry no. 5251; approval date: 21 March 2024; protocol v1.2). The study was conducted in accordance with the Declaration of Helsinki. Given the retrospective, observational design and the use of pseudonymized data, the requirement for written informed consent was waived by the CEIm.

\section*{Data availability statement} The imaging and clinical data used in this study are not publicly available due to patient privacy restrictions under the approved ethics protocol. The code used for the analysis is publicly available at \url{https://github.com/marielatalaveram/radiomics-survival-validation-protocol}.

\section*{Declaration of competing interests}
The authors declare that they have no known competing financial 
interests or personal relationships that could have appeared to 
influence the work reported in this paper.

\section*{CRediT Authorship Contribution Statement}
Mariela Talavera-Mu{\~n}oz: Conceptualization, Formal analysis, Investigation, Methodology, Software, Validation, Visualization, Writing -- original draft, Writing -- review \& editing.
Cecilia Diana-Albelda: Conceptualization, Data curation, Formal analysis, Software, Supervision, Visualization, Writing -- review \& editing.
Miguel Antonio Sampedro-Nu{\~n}ez: Conceptualization, Data curation, Supervision, Writing -- review \& editing.
Victor Rodriguez-Laval: Data curation, Supervision, Writing -- review \& editing.
Jesus Bescos: Funding acquisition, Supervision.
{\'A}lvaro Garc{\'i}a-Mart{\'i}n: Funding acquisition, Supervision, Writing -- review \& editing.

\section*{Declaration of generative AI and AI-assisted technologies in the writing process}
During the preparation of this work, the authors used Claude (Anthropic) to improve language and clarity and to assist with the code used to generate the figures. All figures were produced from the study data with reproducible code, and no generative AI tool was used to create or alter any image. After using this tool, the authors reviewed and edited the content as needed and take full responsibility for the content of the publication.

\bibliographystyle{elsarticle-num} 

\bibliography{cas-refs}

\end{document}